\documentclass[usenatbib]{mn2e}
\usepackage[dvips]{graphicx}
\usepackage{amssymb}
\usepackage{txfonts}
\usepackage{colordvi}

\newcommand{\g}{$\gamma$}

\newbox\grsign \setbox\grsign=\hbox{$>$} \newdimen\grdimen \grdimen=\ht\grsign
\newbox\simlessbox \newbox\simgreatbox \newbox\simpropbox
\setbox\simgreatbox=\hbox{\raise.5ex\hbox{$>$}\llap
     {\lower.5ex\hbox{$\sim$}}}\ht1=\grdimen\dp1=0pt
\setbox\simlessbox=\hbox{\raise.5ex\hbox{$<$}\llap
     {\lower.5ex\hbox{$\sim$}}}\ht2=\grdimen\dp2=0pt
\setbox\simpropbox=\hbox{\raise.5ex\hbox{$\propto$}\llap
     {\lower.5ex\hbox{$\sim$}}}\ht2=\grdimen\dp2=0pt

\def\la{\mathrel{\copy\simlessbox}}

\title[The minimum jet power and equipartition]{The minimum jet power and equipartition}

\author[A. A. Zdziarski]
{Andrzej A. Zdziarski\\
Centrum Astronomiczne im.\ M. Kopernika, Bartycka 18, PL-00-716 Warszawa, Poland\\
}

\date{Accepted 2014 September 4. Received 2014 August 2; in original form 2014 June 3}

\pagerange{\pageref{firstpage}--\pageref{lastpage}}
\pubyear{2014}

\begin{document}

\maketitle

\label{firstpage}

\begin{abstract}
We derive the minimum power of jets and their magnetic field strength based on their observed non-thermal synchrotron emission. The correct form of this method takes into account both the internal energy in the jet and the ion rest-mass energy associated with the bulk motion. The latter was neglected in a number of papers, which instead adopted the well-known energy-content minimization method. That method was developed for static sources, for which there is no bulk-motion component of the energy. In the case of electron power-law spectra with index $>$2 in ion-electron jets, the rest-mass component dominates. The minimization method for the jet power taking it into account was considered in some other work, but only based on either an assumption of a constant total synchrotron flux or a fixed range of the Lorentz factors. Instead, we base our method on an observed optically-thin synchrotron spectrum. We find the minimum jet power is independent of its radius when the ion rest-mass power dominates, which becomes the case below certain critical radius. This allows for robust minimum power estimates. We also present results for the case with observed turnover frequency, at which the source becomes optically thick. This method allows a determination of the source size, in addition to the power and the magnetic field. We also point out that when the ion rest-mass power dominates, the estimates of the minimum power lead to very different equipartition parameters than those based on minimization of the energy content. The former and latter lead to approximate equipartition between the internal energy in magnetic field and in particles including and excluding, respectively, their rest mass energy.
\end{abstract}

\section{Introduction}
\label{intro}

Relativistic jets are common in AGNs, accreting black-hole and neutron-star binaries, \g-ray bursts, and occasionally appear in tidal disruption events. Calculating their power is of major importance, in particular for our understanding of the jet-launching mechanisms, which issue is currently intensely debated (e.g., \citealt{diaztrigo13}), and the impact on their environment. Those jets commonly emit synchrotron radiation.

The energy content of a source emitting synchrotron radiation, e.g., a supernova remnant or a radio lobe, is commonly estimated either assuming equipartition between the energy in the relativistic electrons, $W_{\rm e}$, and in the magnetic field, $W_B$, or by calculating the minimum of $W_{\rm e}+ W_B$ given the observed synchrotron flux, which also results in an approximate equipartition, $W_{\rm e}\sim W_B$ \citep{burbidge56,pacholczyk70}. The corresponding magnetic field strength, $B$, can be calculated as well. This method has been used, e.g., for calculating the minimum energy stored in (static) radio lobes of extragalactic radio sources, at which their jets terminate. When measured, the electron energy is commonly found to be in equipartition with that of the magnetic field, see, e.g., \citet{stawarz13}. The minimum (or measured) energy can then be divided by an estimated jet life time to yield an estimate of the jet power \citep{willott99}.

The same calculation has also been performed to calculate the internal energy content {\it within\/} jets, e.g., by \citet*{bnp13}, hereafter BNP13. Furthermore, the jet power has often been estimated as $(W_{\rm e}+ W_B)/\Delta t$ (e.g., \citealt*{fender99,fbg04,fender06,millerjones06,fgr10,rgf13,brocksopp13}), where $\Delta t\simeq r_{\rm j}/v_{\rm j}$ is a characteristic time scale, e.g., the rise time of a flare and $r_{\rm j}$ and $v_{\rm j}$ is the jet local radius and bulk velocity, respectively. However, this procedure appears incorrect if ions are present in the flow. The jet power is calculated in the frame of the system launching the jet and counterjet, e.g., a black hole surrounded by an accretion disc. In that frame (hereafter called the system frame), the jet energy consists of both its internal energy {\it and\/} the energy of the bulk motion. The jet power equals the enthalpy density flux integrated over the flow cross section (e.g., \citealt{levinson06}). In a simple case of relativistic electrons, cold ions, and a non-relativistic bulk velocity, the jet power in particles equals
\begin{equation}
P_{\rm p}\simeq \left({4 u_{\rm p}\over 3} +{\rho v_{\rm j}^2\over 2}\right)2\upi r_{\rm j}^2 v_{\rm j},
\label{power_nr}
\end{equation}
where $u_{\rm p}$ is the internal energy density in particles, $\rho$ is the mass density, the cross section is assumed to be circular, and both the jet and counterjet are taken into account. The relativistic form of this equation is given in Section \ref{power}. We note that a correction factor accounting for the energy of ions, commonly introduced, includes only their internal energy in the comoving frame, but it still does not account for the bulk motion energy of the ions (either cold or hot) in the system frame.

The neglect of the term associated with the bulk motion can result in a major underestimate of the actual power in particles. In a plasma containing cold protons and relativistic electrons, the relativistic particle power associated with the internal electron energy needs to be multiplied by the factor,
\begin{equation}
A = 1+{m_{\rm p}(\Gamma_{\rm j}-1)\over (4/3)m_{\rm e}\langle\gamma-1\rangle\Gamma_{\rm j}},
\label{correction}
\end{equation}
where $m_{\rm p}$ and $m_{\rm e}$ are the proton and electron mass, respectively, $\langle\gamma\rangle$ is the average Lorentz factor of electrons in the comoving frame, and $\Gamma_{\rm j}$ is the Lorentz factor of the bulk motion corresponding to the velocity of $v_{\rm j}=\beta_{\rm j}c$. For power-law electrons with an index $p>2$ (which is typical for many acceleration processes) and the minimum Lorentz factor of $\gamma_{\rm min}$, we have $\langle\gamma \rangle \simeq  \gamma_{\rm min}(p-1)/(p-2)$, while $\gamma_{\rm min}$ is often $\ll m_{\rm p}/m_{\rm e}$. Thus, this correction factor can be (in some cases) as high as $\sim m_{\rm p}/m_{\rm e}$.

On the other hand, if the energy content is minimized and the contribution due to the bulk motion of ions properly added afterwards (e.g., \citealt{fp00}), the minimum jet power obtained in this way is {\it overestimated}. The reason for it is that the minimum of the comoving energy content occurs at a lower value of the magnetic field than that corresponding to the minimum of the power, thus away from the actual minimum. This effect is related to equipartition of the internal energy corresponding to significantly different parameters than equipartition of the jet power, which was pointed out by \citet{ghisellini99}. 

Minimization of the jet power has been done by \citet{gliozzi99}, \citet{gc01} and \citet{stawarz04} by assuming the total synchrotron power is known, and by \citet{da04} and \citet{dm09} by assuming the known minimum and maximum Lorentz factors in the electron distribution. However, neither of those quantities are usually determined by observations. 

Furthermore, those works simply added the rest mass energy to the internal energy. This corresponds to replacing the (non-relativistic) $\rho v^2/2$ term in equation (\ref{power_nr}) by $\rho c^2\Gamma_{\rm j}$ instead of the correct form of $\rho c^2(\Gamma_{\rm j}-1)$. This is a good approximation for $\Gamma_{\rm j}\gg 1$, but obviously it introduces a large error in systems with low estimated values of $\Gamma_{\rm j}$, e.g., 1.25 for Cyg X-1 \citep{stirling01,gleissner04}. Also, those works neglected the pressure contribution to the power.

We note that the knowledge of the flux from either the synchrotron self-Compton or the external Compton processes (in the cases with the external seed flux known) will determine the magnetic field strength. Then no minimization or equipartition assumptions are needed. However, often no such information is available, and then the presented method gives a stringent lower limit on the jet power.

Here, we calculate the minimum jet power based directly on measured synchrotron fluxes in a certain range. This is the most conservative approach, analogous to the classical method for minimization of the internal energy of \citet{pacholczyk70}. Since the jet power in particles consists of two separate terms, corresponding to their internal energy flux and the rest-mass bulk motion, the resulting equations differ from those for the internal energy minimization. We present results first for the case of the observed synchrotron emission being entirely optically thin. The minimum power depends then on the source size; however, we find a remarkable result that it is independent of the size if the jet particle power is dominated by the ion rest mass. Then, we consider the case in which we observe the spectral turnover corresponding to the onset of synchrotron self-absorption (which method for the energy-content minimization was introduced by \citealt{sr77}). In this case, the source size corresponding to the minimum power is a derived quantity. We consider an emitting source to be clumpy and either be a uniform sphere or  have spherical symmetry with Gaussian distribution of the energy densities (Appendix \ref{gauss}). Finally, we consider steady-state extended jets emitting synchrotron radiation along their length in partially self-absorbed regime.

\section{Definitions}
\label{def}

\subsection{The electron distribution and the synchrotron process}
\label{syn}

In this section, all the quantities are calculated in the comoving (jet) frame. For notational simplicity, we follow the usual convention of using a prime for this frame only for the quantities that are measured in more than one frame, namely the luminosity, photon energy and volume. 

It is possible (and likely) that jets (or synchrotron-emitting blobs) are clumpy. If the spatial distribution of relativistic electrons is clumpy but the magnetic field is uniform, this would have no effect on the minimum power or energy content, since these quantities for synchrotron-emitting particles depend on the total number of particles regardless of their density distribution. However, we expect the magnetic field strength to be coupled to the density of the relativistic electrons (as indeed usually assumed, e.g., \citealt{stawarz04}). This is to a region with a stronger tangled magnetic field to be able to capture more electrons than one with weaker magnetic field. Here, we consider a simple model in which the clumps occupy a fraction, $f$, of the comoving source volume, $V'$, and the space between clumps is devoid of both particles and magnetic field. The probability to be within a clump at a given point is then equal to $f$. The limit of $f=1$ corresponds to a uniform emitting source.

The clumps contain relativistic electrons and (possibly) positrons, which steady-state distribution is given by $N(\gamma)$. We assume $N(\gamma)$ to be a power-law,
\begin{equation}
N(\gamma)=\cases{
K \gamma^{-p}, &$\gamma_{\rm min}\leq \gamma\leq \gamma_{\rm max}$;\cr
0,&otherwise,}
\label{N_el}
\end{equation}
where $K$ is the normalization. The total density of electrons in this distribution and the energy density in all particles (excluding the rest mass) are
\begin{equation}
n_{\rm pl}={K\over p-1}\left(\gamma_{\rm min}^{1-p}-\gamma_{\rm max}^{1-p}\right), \quad u_{\rm p} =n_{\rm pl}m_{\rm e}c^2 \langle\gamma -1\rangle(1+k_{\rm u}),
\label{npl}
\end{equation}
respectively, where $k_{\rm u}$ is the ratio of the energy density of electrons outside the power law and from energetic ions to that in the power-law electrons. The average power-law Lorentz factor and its square are, respectively,
\begin{equation}
\langle\gamma\rangle={p-1\over p-2}{\gamma_{\rm min}^{2-p}-\gamma_{\rm max}^{2-p}\over \gamma_{\rm min}^{1-p}-\gamma_{\rm max}^{1-p}},\quad
\langle\gamma^2\rangle={p-1\over 3-p}{\gamma_{\rm max}^{3-p}-\gamma_{\rm min}^{3-p}\over \gamma_{\rm min}^{1-p}-\gamma_{\rm max}^{1-p}}.\label{av_gamma}
\end{equation} 
When either $p=1$, 2, 3, equations (\ref{npl}--\ref{av_gamma}) need to be trivially modified by using $\ln(\gamma_{\rm max}/\gamma_{\rm min})$. Then, we denote the particle pressure, the density of all positrons (if present; both in the power-law distribution and outside it), and the density of all electrons and positrons as $p_{\rm p}$, $n_+$, and $n_{\rm e}$, respectively. 

We define a dimensionless jet-frame photon energy, $\epsilon'\equiv h\nu'/ (m_{\rm e}c^2)$, and use an approximate relation between $\epsilon'$ and $\gamma$, $\epsilon'\simeq  \gamma^2 (B/B_{\rm cr})$, where $h$ is the Planck constant, $B_{\rm cr}={2\upi m_{\rm e}^2 c^3/(e h)}$ is the critical magnetic field, and $e$ is the electron charge. The optically-thin synchrotron luminosity per $\epsilon'$ of power-law electrons with $\gamma\gg 1$ from volume, $V'$ (e.g., at some height along a jet), containing tangled magnetic field is
\begin{equation}
L'_{\epsilon'}\equiv {{\rm d}L'_{\rm S}\over {\rm d}\epsilon'}\simeq
{C_1(p) \sigma_{\rm T}c K B_{\rm cr}^2 f V' \over 12\upi}\left(B\over B_{\rm cr}\right)^{{p+1}\over 2} 
(\epsilon')^{{1-p}\over 2}.
\label{synspecpl}
\end{equation}
where $\sigma_{\rm T}$ is the Thomson cross section, $C_1(p)$ is given by (e.g., \citealt*{jos74}),
\begin{equation}
C_1(p) = {3^{p+4\over 2} \Gamma\left(3 p-1\over 12\right) \Gamma\left(3 p+19\over 12\right) \Gamma\left(p+1\over 4\right) \over 2^5\upi^{1\over 2}\Gamma\left(p+7\over 4\right)},
\label{c1}
\end{equation}
$\Gamma$ is the gamma function, and the values of $C_1(1.7,\,2,\,2.5,\,3)$ are 1.30, 1.14, 1.01, 1, respectively. Equation (\ref{synspecpl}) allows us to obtain $K$ as a function of the measured quantities, $L'_{\epsilon'}$, $\epsilon'$ and $p$, and the unknown $B$. The comoving volume, $V'$, can be either known or unknown. We then introduce symbols,
\begin{equation}
e_{\rm min}\equiv {\epsilon'_{\rm min}\over \epsilon'},\quad e_{\rm max}\equiv {\epsilon'_{\rm max}\over \epsilon'}.
\label{em}
\end{equation}
Note that these ratios are the same in the jet and observer frames. We assume we know the luminosity of the (approaching) jet, either resolved from the counterjet or dominating the total synchrotron luminosity. We then have in the case of a spherical clumpy source within a clump in the comoving frame,
\begin{eqnarray}
\lefteqn{n_{\rm pl}={9 L'_{\epsilon'}\over C_1(p) \sigma_{\rm T}c B B_{\rm cr} f r_{\rm j}^3(p-1)}\left(e_{\rm min}^{{1-p}\over 2}-e_{\rm max}^{{1-p}\over 2}\right),\label{n_pl}}\\
\lefteqn{u_{\rm p}={9 L'_{\epsilon'} m_{\rm e}c{\epsilon'}^{1\over 2}(1+k_{\rm u})\over C_1(p) \sigma_{\rm T} B^{3\over 2} B_{\rm cr}^{1\over 2} f r_{\rm j}^3(p-2)}\left(e_{\rm min}^{{2-p}\over 2}-e_{\rm max}^{{2-p}\over 2}\right),\label{u_pl}}
\end{eqnarray}
where $e_{\rm min,max}=\gamma_{\rm min,max}^2 B/(\epsilon' B_{\rm cr})$, and $u_{\rm p}$ assumes $\langle\gamma\rangle\gg 1$. In the case of a source with Gaussian energy densities, Appendix \ref{gauss}, the numerical coefficients above should be multiplied by $(p+5)^{3/2}/(6 \upi^{1/2})$. We note that the electron Lorentz factor dominating the emission at $\epsilon'$ has to be within the assumed range of the Lorentz factors,
\begin{equation}
1<\gamma_{\rm min}\la \left(\epsilon' B_{\rm cr}\over B \right)^{1\over 2}\la \gamma_{\rm max}.
\label{eps_min}
\end{equation}
The frequency-integrated synchrotron luminosity (including the case of $\gamma\sim 1$ but neglecting self-absorption) is
\begin{equation}
L'_{\rm S}={B^2\over 6\upi}\sigma_{\rm T} c f V' n_{\rm pl} \langle\gamma^2 -1\rangle.
\label{Lblob}
\end{equation}
The synchrotron self-absorption coefficient for power-law electrons {\it within a clump\/} averaged over the pitch angle can be expressed as (e.g., \citealt{jos74}),
\begin{eqnarray}
\lefteqn{
\alpha_{\rm S}(\epsilon')=  {\upi C_2(p)\sigma_{\rm T} K\over  2\alpha_{\rm f}}  \left(B\over B_{\rm cr}\right)^{p+2\over 2} (\epsilon')^{-{p+4\over 2}},\label{alphas}}\\
\lefteqn{
C_2(p)={3^{p+3\over 2} \Gamma\left(3p+2\over 12\right) \Gamma\left(3p+22\over 12\right) \Gamma\left(p+6\over 4\right)\over 2^4\upi^{1\over 2} \Gamma\left(p+8\over 4\right)},}
\end{eqnarray}
where $\alpha_{\rm f}$ is the fine-structure constant and and the values of $C_2(1.7,\,2,\,2.5,\,3)$ are 0.62, 2/3, 0.80, 1.00, respectively. 

If the source moves, we can relate the flux, $F$, and photon energy, $\epsilon$, in the observer frame to the luminosity and photon energy, respectively, emitted in the jet frame using the Doppler factor, $\delta=1/\left[\Gamma_{\rm j}(1\mp \beta_{\rm j}\cos i)\right]$, where the $-$ and $+$ signs correspond to the jet and counterjet, respectively. The emitted and received frequencies are related by $\epsilon'=\epsilon(1+z)/\delta$, where $z$ is the cosmological redshift. Here, we consider the synchrotron emission from tangled magnetic field and thus to be isotropic in the source frame, and the emitting volume to be moving with the jet (except for Section \ref{extended}). Thus 
\begin{equation}
L'_{\rm S}=\delta^{-4} 4\upi D_L^2 F_{\rm S},\quad L'_{\epsilon'}=\delta^{-3}(1+z)^{-1} 4\upi D_L^2 F_{\epsilon},
\label{lum1}
\end{equation}
where $D_L$ s the luminosity distance and $F_\epsilon$ and $F_{\rm S}$ is the observed synchrotron flux at $\epsilon$ and that integrated over all photon energies, respectively. If the spectrum is a power law with the energy index $\alpha$, $L_{\epsilon} \propto \epsilon^{-\alpha}$, we also have
\begin{equation}
L'_{\epsilon}=\delta^{-3-\alpha}(1+z)^{\alpha-1} 4\upi D_L^2 F_{\epsilon},
\label{lum2}
\end{equation}
where, as often done, we connected the intrinsic luminosity and the observed flux at the same energy.

\subsection{The jet power}
\label{power}

In this section, the mass flow rate, the components of the jet power, and the jet velocity are in the system frame, while the particle and energy densities are in the jet frame. We consider a jet having the local radius $r_{\rm j}$ (the same in both frames). Note that we assume neither a constant jet opening angle nor constant bulk velocity of the jet. The mass flow rate in ions in the jet and counterjet at the considered position along the jet is then
\begin{equation}
\dot M_{\rm j}= 2\upi \mu_{\rm pl} n_{\rm pl} 
m_{\rm p}c \beta_{\rm j}\Gamma_{\rm j} f r_{\rm j}^2, 
\label{mdot}
\end{equation}
where $\mu_{\rm pl}$ is the mean molecular weight per a power-law electron,
\begin{equation}
\mu_{\rm pl}\equiv {\mu_{\rm i}n_{\rm i}\over n_{\rm pl}}, \quad \mu_{\rm i}={4\over 1+3X},
\label{mue}
\end{equation}
$n_{\rm i}$ and $\mu_{\rm i}$ are the ion density and the mean molecular weight, respectively, and $X$ is the hydrogen mass fraction ($\simeq 0.7$ for the standard cosmic composition). For an electron-ion plasma, $\mu_{\rm pl}\geq \mu_{\rm e}$, where $\mu_{\rm e}=2/(1+X)$ is the electron mean molecular weight, the equality sign above corresponds to the absence of cold electrons, and $\mu_{\rm e}=1$ for protons being the only ions. 

We divide the power in particles into the part that can be fully dissipated, which we denote by $P_{\rm e}$ (although it may contain a contribution from hot ions), and the part in the ion rest mass, denoted $P_{\rm i}$, in which case only the bulk-motion kinetic power can be dissipated. The power in the magnetic field is denoted by $P_B$. The power components of the jet and counterjet are then given by 
\begin{eqnarray}
\lefteqn{P_{\rm e}=2\upi  \left(u_{\rm p}+p_{\rm p}+2 n_+ m_{\rm e} c^2\right) c\beta_{\rm j}f (\Gamma_{\rm j} r_{\rm j})^2\nonumber}\\
\lefteqn{\quad \,
= 2\upi  \left[\eta n_{\rm pl}\langle\gamma-1\rangle (1+k)+2 n_+ \right] m_{\rm e}c^3 \beta_{\rm j} f(\Gamma_{\rm j}r_{\rm j})^2,\label{pe}}\\
\lefteqn{P_{\rm i} = \dot M_{\rm j} c^2(\Gamma_{\rm j}-1)=2\upi \mu_{\rm pl} n_{\rm pl} 
m_{\rm p}c^3 \beta_{\rm j}\Gamma_{\rm j} (\Gamma_{\rm j}-1)f r_{\rm j}^2, \label{pi}}\\
\lefteqn{P_B=2\upi (u_B+p_B) \beta_{\rm j} c f(\Gamma_{\rm j} r_{\rm j})^2
= \eta_B {B^2\over 4} c\beta_{\rm j}f (\Gamma_{\rm j} r_{\rm j})^2.\label{pb}}
\end{eqnarray}
Here all the enthalpies are within clumps only, $4/3\leq \eta \leq 5/3$ is the average adiabatic index, $k\geq 0$ is the ratio of the contribution to the proper enthalpy (excluding the rest mass) from electrons outside the power law distribution and from hot ions to that in the power-law distribution, $u_B$ and $p_B$ are the comoving magnetic energy density and pressure (within the clumps), respectively, and $4/3\leq \eta_B \leq 2$ is the magnetic adiabatic index, which range spans the cases from fully tangled to fully ordered toroidal magnetic field \citep{leahy91}. If the magnetic field is ordered but not purely toroidal, the above expressions for $P_B$ need to be multiplied by $(1-\cos\chi)$, where $\chi$ is the angle between the direction between the ordered field and the flow direction (see, e.g., \citealt{meier12}). Note that $k\neq k_{\rm u}$ as the equations of state for the electrons and ions will be, in general, different. We also denote the jet power in particles as $P_{\rm p}=P_{\rm e}+P_{\rm i}$, and the total jet power as $P_{\rm j}$. In the case of a pure power-law pair plasma, $k=0$, $n_{\rm pl}=2 n_+$ and $\mu_{\rm pl}=0$, and then
\begin{equation}
P_{\rm e}=2\upi  \eta n_{\rm pl} \langle\gamma\rangle  m_{\rm e}c^3 \beta_{\rm j} f(\Gamma_{\rm j} r_{\rm j})^2,\quad P_{\rm i}=0,
\label{pairs}
\end{equation}
and the corresponding mass flow rate is
\begin{equation}
\dot M_{\rm j}=2\upi  n_{\rm pl} m_{\rm e}c \beta_{\rm j} \Gamma_{\rm j}f r_{\rm j}^2.
\label{mdot_pair}
\end{equation}

Equation (\ref{pi}) gives the energy flux of the ion rest mass, which is equal to the velocity, $\beta_{\rm j}c$, times the kinetic energy per unit length in the system frame, which is the mass density times the cross-sectional area $\times (\Gamma_{\rm j}-1)c^2$. The additional factor of $\Gamma_{\rm j}$ is then due to the transformation from the jet-frame ion density. We include the e$^\pm$ rest energy in the definition of $P_{\rm e}$ since it had to be expended during production of the pairs, as well as can be recovered when the pairs eventually annihilate. The jet power is defined using the proper enthalpy flux instead of the internal energy flux, which follows from the form of the stress-energy tensor (e.g., \citealt{levinson06}). A physical interpretation of it is that the enthalpy includes both the internal energy and the work done during the jet expansion. Thus, the total jet power, consisting of the above components, had to be supplied to the jet by its formation mechanism, as well as it can be converted to heat and radiation when the jet disappears completely. Also, we include here both the jet and counterjet, since the net momentum flux in the outflows should be null.

As we noted in Section \ref{intro}, our definitions of the jet power differ from many other ones found in the literature. For example, \citet{gliozzi99}, \citet{gc01} and \citet{da04} defined the ion rest-mass jet power with the $\Gamma_{\rm j}$ factor instead of $(\Gamma_{\rm j}-1)$. Also, they used the internal energy instead of the enthalpy for all components, and neglected the counterjet. As discussed above, this appears inaccurate. 

\subsection{Equipartition parameters}
\label{equipartition}

The most common definition of the parameter describing the degree of equipartition between particles and magnetic field uses their pressure ratio, $p_{\rm p}/ p_B$. On the other hand, the well-known method of minimizing the energy content, see Section \ref{min_W} below, usually takes into account the energy density (excluding rest energy), not pressure, while the relationship between magnetic energy density and pressure depends on the geometry of the field, often unknown. Thus, we define here the equipartition parameter in terms of the energy densities (without the particle rest energy),
\begin{equation}
\beta={u_{\rm p}\over u_B}={n_{\rm pl}m_{\rm e}c^2 \langle\gamma-1\rangle(1+k_{\rm u})\over u_B}. 
\label{el_eq}
\end{equation}
For highly relativistic electrons, $k_{\rm u}=0$, and tangled magnetic field, we have $\beta=p_{\rm p}/ p_B$. For $\beta=1$, $P_{\rm e}\simeq P_B$.

On the other hand, the magnetization parameter is defined in terms of the proper enthalpies of the components including the rest energy,
\begin{equation}
\sigma\simeq {u_B+ p_B\over u_{\rm p}+p_{\rm p}+ \mu_{\rm i}n_{\rm i}m_{\rm p}c^2 +n_{\rm e} m_{\rm e}c^2}, 
\label{magn}
\end{equation}
where the last term in the denominator is negligible except in the case of pair-dominated plasma. The reason for including the rest energy here is that the flow inertia is a main factor determining the dynamical behaviour of a system. Since $\sigma$ has the magnetic term in the numerator, we will hereafter compare it with $\beta^{-1}$.

We notice that the magnetization parameter is almost equal to the ratio of the jet magnetic to particle power. The magnetization parameter corresponding to a given jet-power ratio for an electron-ion plasma is 
\begin{eqnarray}
\lefteqn{
\sigma={P_B\over P_{\rm e}+P_{\rm i}} {\mu_{\rm pl} m_{\rm p}(\Gamma_{\rm j}-1)+\eta\langle\gamma-1\rangle (1+k)m_{\rm e}\Gamma_{\rm j}\over \mu_{\rm pl}
m_{\rm p}\Gamma_{\rm j}+\eta\langle\gamma-1\rangle(1+k) m_{\rm e}\Gamma_{\rm j}} 
\nonumber}\\
\lefteqn{
\quad\simeq \cases{\displaystyle{ {P_B\over P_{\rm e}+P_{\rm i}},} & if $\Gamma_{\rm j}\gg 1$;\cr
\displaystyle{{P_B\over P_{\rm e}+P_{\rm i}} {\Gamma_{\rm j}-1\over \Gamma_{\rm j}},} & if $P_{\rm i}\gg P_{\rm e}$.}
\label{sigma_min}}
\end{eqnarray}
Thus, a given ratio of $P_B/(P_{\rm e}+P_{\rm i})$ corresponds to a similar value of the magnetization parameter, except for accounting for the factor $\Gamma_{\rm j}(\Gamma_{\rm j}-1)$ and $\Gamma_{\rm j}^2$ appearing in the respective definitions of $P_{\rm i}$ and $P_{\rm e}$, $P_{\rm B}$. 

We stress that the condition of equipartition may have very different meaning depending which of the two parameters is used (as earlier pointed out by \citealt{ghisellini99}, who considered equipartition of the jet powers rather than the magnetization parameter). If $p>2$, which is usual for many classes of sources, $\langle \gamma \rangle \sim \gamma_{\rm min}$, which is usually $\ll m_{\rm p}/m_{\rm e}$, which in turn implies $\beta^{-1}\gg \sigma$. For example, the model with $p=3.2$ (after cooling) and $\gamma_{\rm min}=2$ of \citet*{mzc13} has $\beta^{-1}\simeq 1.6\times 10^{-4}$ but $\sigma\la 3.2\times 10^{-7}$, see Table 2 of \citet{z14b}. Similarly, the model with $p=2.5$ (of the accelerated electrons, before cooling) and $\gamma_{\rm min}=2$ of \citet{z14b} has $\beta^{-1}\simeq 6.3\times 10^{-2}$ but $\sigma\la 2.0\times 10^{-4}$. Thus, the magnetization parameter in those models is by $\sim$2--3 orders of magnitude lower than the electron equipartition parameter, $\beta^{-1}$, even without accounting for the presence of any ions associated with the likely presence of cold electrons. That presence would increase even more the contrast between the values of these two parameters. 

\section{The minimization method}
\label{min}

\subsection{The minimum energy content}
\label{min_W}

Although we are concerned here with estimating the minimum jet power, we briefly review the case of the energy content, $W$, to enable us to see similarities and differences between the two methods. \citet{burbidge56} was the first to notice that the energy content (in the electrons and magnetic field) of a synchrotron-emitting source is minimized close to equipartition between the energy content in the two components, $\beta\sim 1$. This method has been used in many papers, with its first quantitative description being that of \citet{pacholczyk70}. The results in this section do not depend on the pair abundance.

If we consider a given synchrotron flux integrated over all emitted frequencies, equation (\ref{Lblob}), we find the energy content in particles of $W_{\rm p}=u_{\rm p}f V'=a_{\rm p} L'_{\rm S}B^{-2}$, where the constant $a_{\rm p}$ can be easily obtained from equations (\ref{npl}--\ref{av_gamma}), (\ref{Lblob}), and $W_B=(B^2/8\upi)f V'$. By differentiating $W_{\rm p}+W_B$ over $B$ we readily find that the energy equipartition parameter, the corresponding magnetic field and the minimum total energy content are
\begin{equation}
\beta=1, \quad B_{\rm min}=\left(8\upi a_{\rm p}L'_{\rm S}\over f V'\right)^{1\over 4},
\quad W_{\rm min}=\left(a_{\rm p} L'_{\rm S}f V' \over 2\upi\right)^{1\over 2},
\label{Wmin1}
\end{equation}
respectively. This holds independently of the electron distribution, even if it is different from a power law. Such a source has the minimum energy content at a given $L'_{\rm S}$, or, equivalently, is the most efficient synchrotron emitter at a given energy content. Note that hereafter $B_{\rm min}$ denotes $B$ corresponding to $W_{\rm min}$ rather than a minimum magnetic field strength. 

If we then know the synchrotron flux at some energy, $L'_{\epsilon'}$, we can use it to derive the minimum energy content. We first assume the range of the electron Lorentz factors, $\gamma_{\rm min}$ and $\gamma_{\rm max}$, e.g., from knowing that the electrons are accelerated up to some Lorentz factor and then cooled down to a another one. Equations (\ref{npl}--\ref{synspecpl}) imply $W_{\rm p}=u_{\rm p}f V'= a_{\rm p} L'_{\epsilon'} B^{-(p+1)/2}$, where, for notational simplicity, we hereafter use the same symbol $a_{\rm p}$ in various local contexts. Then, proceeding analogously as above, we find
\begin{eqnarray}
\lefteqn{
\beta^{-1}={p+1\over 4},\quad B_{\rm min}=\left[2 \upi a_{\rm p}L'_{\epsilon'} (p+1) \over f V'\right]^{2\over 5+p},\nonumber}\\
\lefteqn{
W_{\rm min}={p+5\over 4}\left(a_{\rm p}L'_{\epsilon'}\right)^{4\over 5+p}\left[f V'\over 2\upi (p+1)\right]^{1+p\over 5+p}. 
\label{Wmin2}}
\end{eqnarray}

On the other hand, we can instead assume, as usually done (e.g., \citealt{pacholczyk70,longair11,willott99}; BNP13) that we know $\epsilon'_{\rm min}$ and $\epsilon'_{\rm max}$. This yields, using equations (\ref{synspecpl}--\ref{u_pl}), $\gamma_{\rm min,max}=(\epsilon'_{\rm min,max} B_{\rm cr}/B)^{1/2}$, which implies $W_{\rm p}=u_{\rm p}f V'=a_{\rm p}L'_{\epsilon'} B^{-3/2}$ (where, as before, the definition of $a_{\rm p}$ is local). This implies
\begin{equation}
\beta^{-1}={3\over 4},\quad B_{\rm min}=\left(6\upi a_{\rm p}L'_{\epsilon'}\over f V'\right)^{2\over 7},\quad
W_{\rm min}={7\over 4}\left(a_{\rm p}L'_{\epsilon'}\right)^{4\over 7}\left(f V'\over 6\upi\right)^{3\over 7}.
\label{Wmin3}
\end{equation}
This, commonly used, result assumes that there is no synchrotron flux outside $\epsilon'_{\rm min}$--$\epsilon'_{\rm max}$. We often do not know whether this is satisfied, and, if there is such emission, the actual energy content can be higher. Still, this approach gives us the absolute minimum of $W$, while the former gives the minimum corresponding to some physically guessed values of $\gamma_{\rm min}$ and $\gamma_{\rm max}$. We note that although the value of $\beta^{-1}=3/4$ does not depend on $p$, this result does assume that $N(\gamma)$ is a power law, and the coefficients in $B_{\rm min}$ and $W_{\rm min}$ depend on $p$. Also, the last two methods assume that the electrons are relativistic on average, implying that equation (\ref{synspecpl}) applies and the $-1$ term in the definition of $u_{\rm p}$, equation (\ref{npl}), can be neglected. This requires $\gamma_{\rm max}\gg 1$ for $p<2$ (which is usually satisfied), or, more constraining, $\gamma_{\rm min}\gg 1$ for $p>2$. If this is not satisfied, more complex formulae will be obtained. 

\subsection{The minimum jet power}
\label{min_P}

Here, we estimate the jet power in the case its synchrotron emission is localized. The emission is from some moving volume while the instantaneous power is related to the jet cross section. We consider a simple model in which the synchrotron emission comes from a moving volume spherical in the jet frame and we measure its power at its maximum corresponding to the cross section intersecting the sphere centre, with the area $\upi r_{\rm j}^2$. From the momentum conservation, we expect a similar emitting blob to be sent in the opposite direction (although some asymmetry is possible, \citealt{aa99}). This scenario corresponds to transient jets, in which individual blobs are ejected from the central compact object. Such event are often observed, see, e.g.,  \citet{mr94} and \citet{fender99} for the case of blob ejections from stellar-mass binary systems. We consider clumpy spherical sources, in which the clumps are distributed uniformly throughout the source. The source comoving volume is $V'=(4\upi/3) r_{\rm j}^3$, and the clumps occupy a fraction, $f$, of it. In Appendix \ref{gauss}, we also consider sources with Gaussian profiles, which geometry results in a change of some numerical constants. We proceed analogously to the treatment in Section \ref{min_W}. 

We first consider the frequency-integrated synchrotron luminosity. In this case, equations (\ref{npl}--\ref{av_gamma}), (\ref{Lblob}) give $n_{\rm pl}\propto L'_{\rm S}(f V')^{-2} B^{-2}$, which can be substituted in equations (\ref{pe}--\ref{pi}) to describe maximally synchrotron-emitting jets. Then, using equations (\ref{pe}--\ref{pb}), the jet power in particles can be written as $P_{\rm p}=a_{\rm p} L'_{\rm S} r_{\rm j}^{-1}B^{-2}$, and that in magnetic fields as $P_B=a_B f r_{\rm j}^2 B^{2}$ (with local definitions of $a_{\rm p}$, $a_B$). From ${\rm d}P_{\rm j}/{\rm d}B=0$, we find the jet power ratio, the corresponding $B$, and the minimum total power at
\begin{equation}
{P_B\over P_{\rm e}+P_{\rm i}}= 1,\quad B_{\rm min}=\left({a_{\rm p} L'_{\rm S}\over a_B f r_{\rm j}^3}\right)^{1\over 4},\quad P_{\rm min}=2\left( a_{\rm p} a_B L'_{\rm S} f r_{\rm j}\right)^{1\over 2},
\label{Pmin1}
\end{equation}
respectively. This approach, using an assumed value of the total $L'_{\rm S}$, was adopted by \citet{gliozzi99}, \citet{gc01} and \citet{stawarz04}. \citet{stawarz04} also considered clumpiness, and first obtained the dependencies on $f$ given in equation (\ref{Pmin1}).

Note that usually we do not know the synchrotron flux integrated over all frequencies, and that the range of the emitted fluxes depends on $B$. Therefore, the above approach is not accurate when the observable is the synchrotron flux at some energy and either the range of observed photon energies or the range the electron Lorentz factors in $N(\gamma)$. If we know $\gamma_{\rm min}$ and $\gamma_{\rm max}$, we have, using equations (\ref{npl}--\ref{synspecpl}) and (\ref{pe}--\ref{pi}), $P_{\rm p}=a_{\rm p} L'_{\epsilon'} r_{\rm j}^{-1} B^{-(p+1)/ 2}$ and $P_B$ as above. From ${\rm d}P_{\rm j}/{\rm d}B=0$, we find \begin{eqnarray}
\lefteqn{
{P_B\over P_{\rm e}+P_{\rm i}}= {1+p\over 4},\quad B_{\rm min}=\left({p+1\over 4}{a_{\rm p}L'_{\epsilon'}\over a_B f r_{\rm j}^3}\right)^{2\over 5+p},\nonumber}\\
\lefteqn{
P_{\rm min}=(p+5)\left(a_{\rm p}L'_{\epsilon'}\over 4\right)^{4\over 5+p}\left(a_B f\over 1+p\right)^{1+p\over 5+p} r_{\rm j}^{2 p-2\over 5+p}.
\label{Pmin2}}
\end{eqnarray}
This approach was adopted by \citet{da04} and \citet{dm09}. The condition (\ref{eps_min}) also applies here. This method involves a guess about the range of the Lorentz factors, which in most cases cannot be determined with certainty. 

On the other hand, we can minimize the jet power yielding only the emission observed at $\epsilon'_{\rm min}$--$\epsilon'_{\rm max}$, analogously to the classical treatment of the minimum energy content. Since $\gamma_{\rm min,max}=(\epsilon'_{\rm min,max} B_{\rm cr}/B)^{1/2}$, equations (\ref{npl}--\ref{synspecpl}) and (\ref{pe}--\ref{pi}) imply $P_{\rm e}=a_{\rm e} L'_{\epsilon'} r_{\rm j}^{-1} B^{-3/ 2}$, $P_{\rm i}=a_{\rm i} L'_{\epsilon'} r_{\rm j}^{-1} B^{-1}$, while $P_B=a_B f r_{\rm j}^2 B^{2}$ as before. The explicit form of these coefficients in the present case is
\begin{eqnarray}
\lefteqn{a_{\rm i}={18\upi m_{\rm p}c^2\mu_{\rm pl}\beta_{\rm j}\Gamma_{\rm j}(\Gamma_{\rm j}-1)\over C_1(p) \sigma_{\rm T} B_{\rm cr} (p-1)}\left(e_{\rm min}^{{1-p}\over 2}-e_{\rm max}^{{1-p}\over 2}\right),\label{a_i}}\\
\lefteqn{a_{\rm e}={24\upi m_{\rm e}c^2\epsilon^{1\over 2}\beta_{\rm j}\Gamma_{\rm j}^2\over C_1(p) \sigma_{\rm T} B_{\rm cr}^{1\over 2} (p-2)}\left(e_{\rm min}^{{2-p}\over 2}-e_{\rm max}^{{2-p}\over 2}\right).\label{a_e}}\\
\lefteqn{a_B={c\beta_{\rm j}\Gamma_{\rm j}^2\over 3},\label{a_B}}
\end{eqnarray}
where $\eta(1+k)=4/3$ (i.e., this component of the power dominated by relativistic power law electrons) was assumed for $a_{\rm e}$, and tangled magnetic field was assumed for $a_B$. In the case of a Gaussian source, the values of $a_{\rm i}$ and $a_{\rm e}$ given above should be multiplied by $(p+5)^{3/2}/(6 \upi^{1/2})$. From ${\rm d}P_{\rm j}/{\rm d}B=0$, we find
\begin{equation}
{3\over 4}P_{\rm e}+{1\over 2}P_{\rm i}=P_B.
\label{Pmin3}
\end{equation}
This equation can be numerically solved for $B$, and then $P_{\rm j}$ can be calculated from equations (\ref{pe}--\ref{pb}). In the limiting case of $P_{\rm e}\gg P_{\rm i}$ we have,
\begin{eqnarray}
\lefteqn{
{P_B\over P_{\rm e}}= {3\over 4},\quad
B_{\rm min}=\left(3 a_{\rm e}L'_{\epsilon'}\over 4 a_B f r_{\rm j}^3 \right)^{2\over 7},\quad P_{\rm min}=7 \left( a_B f\over 3\right)^{3/7} \left( a_{\rm e} L'_{\epsilon'}\over 4\right)^{4\over 7} r_{\rm j}^{2\over 7}.\nonumber}\\
\lefteqn{
\label{Pmin3e}}
\end{eqnarray}
This case also corresponds to pair-dominated jets. Formulae similar to those in equation (\ref{Pmin3e}) have been also obtained by \citet{dm09} by setting $p=2$ in the case described in equation (\ref{Pmin2}). However, our treatment here is valid for any $p$. Then, for $P_{\rm i}\gg P_{\rm e}$ we have,
\begin{equation}
{P_B\over P_{\rm i}}= {1\over 2},\quad
B_{\rm min}=\left(a_{\rm i}L'_{\epsilon'}\over 2 a_B f r_{\rm j}^3 \right)^{1\over 3},\quad P_{\rm min}=3 (a_B f)^{1\over 3}\! \left( a_{\rm i} L'_{\epsilon'}\over 2\right)^{2\over 3}.
\label{Pmin3i}
\end{equation}
If both powers are comparable, the minimum power corresponds to some value of the jet power ratio between 1/2 and 3/4. Remarkably, $P_{\rm min}$ is independent of the source size for $P_{\rm i}\gg P_{\rm e}$. From the above formulae, we find $P_{\rm e}/P_{\rm i}\propto B_{\rm min}^{-1/2}\propto r_{\rm j}^m$, where $m\in [3/7,1/2]$. Thus, the dominance of the ionic power increases with decreasing source size, and the dependence of the minimum power on the size occurs only above some critical value of the radius,
\begin{equation}
r_{\rm cr}\simeq {a_{\rm i}^{7/3} (L'_{\epsilon'})^{1/3}\over a_{\rm e}^2 (a_B f)^{1/3}}.
\label{r_crit}
\end{equation}
If $r_{\rm j}\ll r_{\rm cr}$, the minimum power estimate is independent of $r_{\rm j}$. 

As noted above, the jet kinetic power depends sensitively on its bulk motion, which for the energy content calculations is important only via the Doppler effect. An ionic jet can, for a wide range of its parameters, have its inertia dominated by the rest mass of ions, which effect does not enter the calculations of the minimum energy content. Then a jet with $P_B/(P_{\rm e}+P_{\rm i})\simeq \sigma\simeq 1$ has $\beta^{-1}\gg 1$, i.e., the magnetic energy density is $\gg$ the electron energy density. Vice versa, a jet with $\beta=1$ has $P_{\rm e}=P_B$, which, for $P_{\rm i}\gg P_{\rm e}$, implies $\sigma\ll 1$.

If the matter in the jet consists of a pure e$^\pm$ pair plasma, the results in this section remain except for different definitions of $a_{\rm p}$ or $a_{\rm e}$ and $P_{\rm i}=0$, equation (\ref{pairs}). In particular, equation (\ref{Pmin3e}) applies in the case constrained by emission observed in the $\epsilon'_{\rm min}$--$\epsilon'_{\rm max}$ range. Also, then $\sigma= P_B/P_{\rm e}\simeq \beta^{-1}$.

\subsection{Minimization taking into account self-absorption}
\label{self}

An extension of the energy-content minimization method takes into account synchrotron self-absorption. If the frequency and flux corresponding to the source becoming optically-thick to self-absorption (the turnover) are known, this information can be used in the energy-content minimization method (\citealt{sr77,chevalier98} and many following papers). This method has been, in particular, applied to the energy content of jets in the context of \g-ray bursts (BNP13 and references therein). However, no analogous calculations for the minimum jet power have been done yet.

A synchrotron source may be self-absorbed below certain frequency, see equation (\ref{alphas}). If we know that a frequency, $\epsilon'_{\rm t}$, corresponds to the turnover frequency of the source, i.e., the optical depth to synchrotron self-absorption is $\tau_{\rm S}(\epsilon'_{\rm t})=\alpha_{\rm S}(\epsilon'_{\rm t})f r_{\rm j}=1$. This implies (independent of $f$)
\begin{equation}
B= {a_{\rm t} r_{\rm j}^4\over (L'_{\epsilon'_{\rm t}})^2}\propto {\theta^4 \over F_{\epsilon_{\rm t}}^2},
\label{BrL}
\end{equation}
where $\theta$ is the source angular size, $F_{\epsilon_{\rm t}}$ is the observed flux at the turnover, and $a_{\rm t}$ is given by
\begin{equation}
a_{\rm t}=\left[ 2 C_1(p) \alpha_{\rm f} c \over 9\upi C_2(p)\right]^2 \left(B_{\rm cr} \epsilon_{\rm t}' \right)^5.
\label{a_t}
\end{equation}
For a Gaussian source, Appendix \ref{gauss}, and defining $\tau_{\rm S}=1$ for the line of sight from the source centre, the above value of $a_{\rm t}$ needs to be multiplied by $4(p+6)/(p+5)^3$. If we can measure $\theta$, we need neither the minimum energy nor equipartition argument to calculate the magnetic field strength, which was probably first noticed by \citet{slish63} and \citet{williams63}, and later by, e.g., \citet{hw66}, \citet{bridle67} and \citet{sw68}. In that case, the self-absorption determination of $B$ can be compared with the value of $B$ assuming equipartition, which was probably first done by \citet{bridle67}. 

However, for many sources we know neither their angular nor physical size. We can then substitute the resulting $r_{\rm j}=(L'_{\epsilon'})^{1/2} B^{1/4} a_{\rm t}^{-1/4}$ and estimate both $B$ and $r_{\rm j}$ using either the minimum energy or the equipartition argument. The first such calculation appears to be that of \citet{sr77}, who derived an expression for the angular size of a self-absorbed source assuming equipartition. Then a number of other authors applied this method, e.g., \citet{chevalier98}, BNP13. The method has also been independently re-derived by some other authors, see, e.g., an equipartition energy-content calculation by \citet*{cdr11}. Here, for illustration, we show the well-known result for the minimum energy content given a synchrotron spectrum in the range $\epsilon'_{\rm t}$--$\epsilon'_{\rm max}$. We have the energy content in particles as before, $W_{\rm p}=a_{\rm p}L'_{\epsilon'_{\rm t}} B^{-3/2}$, but now, using the value of $r_{\rm j}$ from equation (\ref{BrL}), $W_B=f a_{\rm t}^{-3/4}(L'_{\epsilon'_{\rm t}})^{3/2} B^{11/4}/6$. This implies
\begin{eqnarray}
\lefteqn{
\beta^{-1}={6\over 11},\quad B_{\rm min}=\left(6^2 a_{\rm p}\over 11 f\right)^{4\over 17} a_{\rm t}^{3\over 17} \left(L'_{\epsilon'_{\rm t}}\right)^{-2\over 17},\label{Wminabs}}\\
\lefteqn{
r_{\rm min}=\left(6^2 a_{\rm p}\over 11 f\right)^{1\over 17} a_{\rm t}^{-7\over 34} \left(L'_{\epsilon'_{\rm t}}\right)^{8\over 17},\quad
W_{\rm min}={17 f^{6\over 17}\over 6^{12\over 17} } 
\left(a_{\rm p}\over 11\right)^{11\over 17} a_{\rm t}^{-9\over 34} \left(L'_{\epsilon'_{\rm t}}\right)^{20\over 17},\nonumber}
\end{eqnarray}
where $r_{\rm min}$ is the radius corresponding to the minimum energy.

We apply an analogous method to determine the minimum jet power and the corresponding size and the magnetic field strength, provided a measured flux is at the turnover energy. Since an optically-thin spectrum is now limited by the measured $\epsilon_{\rm t}$, we consider only the case with the known range of the optically-thin synchrotron emission, $\epsilon'_{\rm t}$--$\epsilon'_{\rm max}$. Substituting equation (\ref{BrL}) in equations (\ref{pe}--\ref{pb}), we find $P_{\rm e}=a_{\rm e} a_{\rm t}^{1/4} (L'_{\epsilon'_{\rm t}})^{1/2} B^{-7/4}$, $P_{\rm i}=a_{\rm i} a_{\rm t}^{1/4} (L'_{\epsilon'_{\rm t}})^{1/2} B^{-5/4}$, $P_B=a_B f a_{\rm t}^{-1/2} L'_{\epsilon'_{\rm t}} B^{5/2}$. Here, $a_{\rm i}$, $a_{\rm e}$ and $a_B$ are given by equations (\ref{a_i}--\ref{a_B}) with $e_{\rm min}=1$. From ${\rm d}P_{\rm j}/{\rm d}B=0$, we find
\begin{equation}
{7\over 10}P_{\rm e}+{1\over 2}P_{\rm i}=P_B.
\label{Pminabs}
\end{equation}
This equation can be numerically solved for $B$. Then $r_{\rm j}$ can be calculated from equation (\ref{BrL}) and $P_{\rm min}$, from equations (\ref{pe}--\ref{pb}). In the limiting case of $P_{\rm e}\gg P_{\rm i}$ we have,
\begin{eqnarray}
\lefteqn{
{P_B\over P_{\rm e}}= {7\over 10},\quad 
B_{\rm min}=\left(7 a_{\rm e}\over 10 a_B f\right)^{4\over 17}\!\!a_{\rm t}^{3\over 17} \left(L'_{\epsilon'_{\rm t}}\right)^{-2\over 17},\label{Pminabse}}\\
\lefteqn{
r_{\rm min}=\left(7 a_{\rm e}\over 10 a_B f\right)^{1\over 17} \!\!\! a_{\rm t}^{-7\over 34}\!\! \left(L'_{\epsilon'_{\rm t}}\right)^{8\over 17}\!\!,\quad
P_{\rm min}=17 a_{\rm t}^{-1\over 17}\!\left( a_B f\over 7\right)^{7\over 17} \!\left( a_{\rm e} \over 10\right)^{10\over 17}\! \left(L'_{\epsilon'_{\rm t}}\right)^{12\over 17}\!\!.\nonumber
}
\end{eqnarray}
The last scaling has also been obtained in \citet{fb95} assuming equipartition between the electron and magnetic energy densities. Then, for $P_{\rm i}\gg P_{\rm e}$ we have,
\begin{eqnarray}
\lefteqn{
{P_B\over P_{\rm i}}= {1\over 2},\quad 
B_{\rm min}=\left(a_{\rm i}\over 2 a_B f\right)^{4\over 15}\!\!a_{\rm t}^{1\over 5} \left(L'_{\epsilon'_{\rm t}}\right)^{-2\over 15},\label{Pminabsi}}\\
\lefteqn{
r_{\rm min}=\left(a_{\rm i}\over 2 a_B f\right)^{1\over 15} \!\!a_{\rm t}^{-1\over 5} \left(L'_{\epsilon'_{\rm t}}\right)^{7\over 15},\quad
P_{\rm min}=3 (a_B f)^{1\over 3} \left( a_{\rm i}L'_{\epsilon'_{\rm t}} \over 2\right)^{2\over 3} .\nonumber
}
\end{eqnarray}
Notably, $P_B/P_{\rm i}$ and $P_{\rm min}$ are the same as in the case without utilizing self-absorption. This is a consequence of the $P_{\rm min}$ in the optically-thin case being independent of $r_{\rm j}$ for $P_{\rm i}\gg P_{\rm e}$. The two above regimes correspond to $L'_{\epsilon'_{\rm t}}$ being $>$ and $<$ than some critical luminosity, respectively. This also corresponds to the source size being $>$ and $<$ than a critical radius, respectively. The critical values are given by,
\begin{equation}
L'_{\rm cr}\simeq {a_{\rm i}^{17} a_{\rm t}^{3/2}\over (a_B f)^2 a_{\rm e}^{15}},\quad r_{\rm cr}\simeq {a_{\rm i}^8 a_{\rm t}^{1/2}\over a_B f a_{\rm e}^7}.
\label{l_crit}
\end{equation}
This critical radius is the same as that given by equation (\ref{r_crit}) after the substitution of $L'_{\rm cr}$.

Interestingly, the minimum power condition provides a rather robust estimate of the source radius (as also noted by BNP13). Since $B\propto r_{\rm j}^4$, equation (\ref{BrL}), the power corresponding to a departure from the minimum condition increases very fast with changing $r_{\rm j}$, especially for an increase of $r_{\rm j}$,
\begin{equation}
P\simeq P_{\rm min}\cases{
\left[{10\over 17}\left(r_{\rm j}\over r_{\rm min}\right)^{-7}+{7\over 17}\left(r_{\rm j}\over r_{\rm min}\right)^{10}\right],& $P_{\rm e}\gg P_{\rm i}$;\cr
\left[{2\over 3}\left(r_{\rm j}\over r_{\rm min}\right)^{-5}+{1\over 3}\left(r_{\rm j}\over r_{\rm min}\right)^{10}\right], & $P_{\rm i}\gg P_{\rm e}$.}
\label{Pvar}
\end{equation}
Thus, a large departure from $r_{\rm min}$ requires a very large power. Analogous expressions can readily be written for other considered cases in terms of powers of $B/B_{\rm min}$.

We note that this method is particularly useful to transient jets, with ejections of blobs. Initially, right after the ejection, the blob is optically thick to self-absorption and its radio luminosity is low. At some point of time, the optical depth reaches unity at the monitored frequency, and the maximum flux is reached. The minimum jet power and the corresponding $B$ and $r_{\rm j}$ should be calculated at this time. Further expansion leads to adiabatic cooling of electrons and decrease of the magnetic field, causing the flux to decline. 

\subsection{Extended jet emission}
\label{extended}

The method presented above assumes that the synchrotron emission is localized, i.e., originates from a well-defined region of the jet, and at least a part of the emission is optically thin. If we observe flat, partially self-absorbed, radio spectra originating in the entire jet, as in the model of \citet{bk79}, we need to modify the method. For a steady-state jet, the relations between the observed isotropic optically-thin flux and the luminosity are given by,
\begin{equation}
L'_{\epsilon'}=\delta^{-2}\Gamma_{\rm j}(1+z)^{-1} 4\upi D_L^2 F_{\epsilon}, \quad L'_{\epsilon}=\delta^{-2-\alpha}\Gamma_{\rm j}(1+z)^{\alpha-1} 4\upi D_L^2 F_{\epsilon}
\label{lum3}
\end{equation}
\citep{sikora97,lb85}. In the model of \citet{bk79}, the jet power is constant along the jet, since both the magnetic field energy flux and the flux in relativistic electrons are conserved. Then, this power can be minimized. 

If we know the turnover energy, we can apply the same method as in Section \ref{self}, because the optically-thin emission of that type of jets is dominated by a region just above the onset of dissipation. A difference with respect to the formalism above is that the emission region is (approximately) cylindrical rather than spherical; its vertical extent is of the order of the height of the dissipation onset, $z_{\rm j}$, and thus the characteristic volume on one side is $\sim \upi r_{\rm j}^2 z_{\rm j}\Gamma_{\rm j}$. Thus, we need to replace $L'_{\epsilon'_{\rm t}}$ by $L'_{\epsilon'_{\rm t}}\Theta_{\rm j}/\Gamma_{\rm j}$ in equations (\ref{Pminabse}--\ref{Pminabsi}), and $L'_{\rm cr}$ by $L'_{\rm cr} \Theta_{\rm j}/\Gamma_{\rm j}$ in equation (\ref{l_crit}), where $\Theta_{\rm j}=r_{\rm j}/z_{\rm j}$ is the jet opening angle of the considered region. In this case, we can solve for the minimum power as a function of the jet opening angle, $\simeq r_{\rm j}/z_{\rm j}$.

On the other hand, we may also consider the case in which we observe only the partially self-absorbed part of the spectrum, with the flux, $F_\epsilon$, and the energy index of $\alpha\simeq 0$ \citep{bk79}, but we do not know the turnover energy. In this case, following \citet{bk79}, we assume a conical jet with a constant bulk velocity and conserved energy fluxes in both magnetic field and electrons. We note that a partially optically-thick synchrotron emission is not isotropic in the jet frame, see, e.g., equations (22--23) of \citet*{zls12}. We can then substitute their equation (23) in their equation (22) to express the normalization of the electron distribution at the jet base, $K_0$, as a function of $F_\epsilon$, the magnetic field, $B_0$, height, $z_0$, at the base, and the opening angle, $\Theta_{\rm j}=r_0/z_0$. Note that since $\alpha=0$, $F_\epsilon$ is constant across the partially self-absorbed part of the spectrum. Here, we neglect possible clumping. Also, since the the extent of this spectrum is not related to the range of $\gamma$ but rather to the ratio of the maximum jet height to that of the base, we cannot readily relate here $\epsilon'$ to $\gamma$. Thus, we assume the values of $\gamma_{\rm min}$ and $\gamma_{\rm max}$, i.e., use equations (\ref{npl}--\ref{av_gamma}) for $n_{\rm pl}$ and $u_{\rm p}$. Then, assuming the factor $2n_+$ is negligible in equation (\ref{pe}), we can write $P_{\rm e}+P_{\rm i}=(a_{\rm e}+a_{\rm i})(F_\epsilon \Theta_{\rm j})^{(p+4)/5}(r_0 B_0)^{-(2 p+3)/5}$ and $P_B=a_B (r_0 B_0)^2$, where $a_B$ is given by equation (\ref{a_B}),
\begin{eqnarray}
\lefteqn{
a_{\rm i}=2\upi \mu_{\rm pl}{n_{\rm pl}\over K}K_0' m_{\rm p}c^3 \beta_{\rm j}\Gamma_{\rm j}(\Gamma_{\rm j}-1),\label{a_extended}}\\
\lefteqn{
a_{\rm e}=2\upi {u_{\rm p}\over K} K_0' c \beta_{\rm j}\Gamma_{\rm j}^2,}\\
\lefteqn{
K_0'={1\over \sigma_{\rm T}B_{\rm cr}}\left(\upi\over \delta\right)^{7+3 p\over 5}\left[C_2(p)\over \alpha_{\rm f} \sin i\right]^{p-1\over 5}\left[24 D_L^2\over c C_1(p)C_3(p)(1+z)\right]^{4+p\over 5}
,}
\end{eqnarray}
$C_3(p)$ is defined in \citet{zls12}, and $K_0'$ is the base electron normalization without dependencies on $F_\epsilon$, $\Theta_{\rm j}$, $B_0$ and $r_0$. Proceeding as before, we find
\begin{eqnarray}
\lefteqn{
{P_B\over P_{\rm e}+P_{\rm i}}= {2p+3\over 10},\quad B_{\rm min}=\left({2 p+3\over 10}{a_{\rm e}+a_{\rm i}\over a_B}\right)^{5\over 2 p+13}\left(F_\epsilon\Theta_{\rm j}\right)^{p +4\over 2 p+13} r_0^{-1} ,\nonumber}\\
\lefteqn{
P_{\rm min}=(2 p+13)\left(a_B\over 2 p+3\right)^{2 p +3\over 2 p+13} \left(a_{\rm e}+a_{\rm i}\over 10\right)^{10\over 2 p+13}\left(F_\epsilon\Theta_{\rm j}\right)^{2 p +8\over 2 p+13},
\label{Pminext}}
\end{eqnarray}
Notably, the minimum jet power is independent of the absolute value of either $z_0$ or $r_0$ and it scales close to linear with both the flux and the jet opening angle, $r_0/z_0$. 

As an example, we apply these results to the hard state of Cyg X-1, which has the radio-mm index of $\alpha\simeq 0$ \citep{fender00}. We use $F_\epsilon$ corresponding to the average radio flux of 15 mJy, $D_L=1.86$ kpc \citep{reid11}, $i=27\degr$ \citep{orosz11}, $\Theta_{\rm j}=2\degr$, which is the upper limit of \citep{stirling01}, and, somewhat arbitrarily, $\beta_{\rm j}=0.6$, $p=2.5$, $\mu_{\rm pl}=1$, $k=0$, $\gamma_{\rm min}=2$, $\gamma_{\rm max}=10^3$ (which value is not important for $p>2$), see, e.g., discussion in \citet{zls12}. We find $P_{\rm min}\simeq 2.7\times 10^{35}$ erg s$^{-1}$ (and $P_{\rm i}\gg P_{\rm e}$), well below the existing estimates, e.g., \citet{gallo05}. If we assume equipartition between the electron and magnetic energy densities, we obtain $P\simeq 8.5\times 10^{35}$ erg s$^{-1}$, also independent of $r_0$. Note that this value is substantially higher than the $P_{\rm min}$, illustrating the difference between the power minimization and $\beta\sim 1$, discussed earlier.

On the other hand, if electron energy losses are taken into account, the jet non-radiative power depends on the height along the jet, e.g., \citet{z14a}. This is because electrons lose energy radiatively, and their energy density is a function of height in a general case, even if we assume acceleration of them along the jet. The electrons also lose energy adiabatically, which energy can be either transferred to an external medium or converted into an increase of the bulk velocity of the jet, see, e.g., discussion in \citet{z14a}.

\section{Discussion}
\label{discussion}

The jet-power minimization method developed here can be applied at many astrophysical settings. It has been applied, in particular, to major outbursts of the black-hole binary GRS 1915+105 in \citet{zdz14}. 

\citet{nm12} and \citet*{smn13} have found a correlation of the maximum radio flux with their estimated black-hole spin in several sources having transient optically-thin jets. Then, as a proxy to the jet power, they used that maximum radio flux. As it is well-known, and also confirmed by the present work, the jet power is not linearly proportional to the radio flux. Still, the two quantities are correlated. If the power of transient jets is dominated by ions, we have found that the minimum power (at the equipartition parameter $\sigma\sim 1$) is independent of the source size and $P_{\rm min}\propto (L'_{\epsilon'})^{2/3}$. This would argue that if the peak fluxes are indeed correlated with the spin, the jet power are also correlated.

On the other hand, \citet{fgr10} and \citet{rgf13} have applied the minimum energy content method to derive the transient jet power, assuming the flare rise time, $\Delta t$, is proportional to the source size, and claimed no such correlation. As our work shows, this method fails to estimate the total jet power if it has a substantial ion component. In particular, the energy content method yields $\beta^{-1}\sim 1$, whereas the minimization of the total jet power corresponds to $\sigma\sim 1$, with $\sigma\gg \beta^{-1}$ under typical circumstances. The energy content method assumes that the electron component dominates the power, which then depends on both $L'_{\epsilon'}$ and the source size (found from the rise time). Since this assumption may be not satisfied, application of that method may introduce a spurious scatter in the derived values of the jet power.

Then, BNP13 have presented a method to constrain the energy content in relativistically moving sources, in particular in \g-ray bursts, by using the minimization of the internal energy in the comoving jet frame and transforming it to the system frame. In addition, they relate the source size and the bulk Lorentz factor to the time since the burst onset assuming a wide jet, which relation they include in their minimization formulae. Furthermore, they account for the energy of hot ions. However, it appears that the method of BNP13 still does not account for the energy associated with the bulk motion of the ion rest mass in the system frame (which contributes even in the absence of any process energizing ions). This can be added using the formalism of the present work. This would yield both the total energy content of a \g-ray burst and its power.

We also note that the minimization method developed in the present work yields, in general, the minimum {\it possible\/} jet power. It is not clear whether the actual jet power may be at that value or even be proportional to the minimum power for a sample of objects. For example, adiabatic and radiative cooling of transient jets usually leads to a fast decrease of their synchrotron luminosity while the kinetic power in ions may remain unaffected. This may have been the case for the 1994 outburst of GRS 1915+105 \citep{mr94}. Then the application of the minimization method at a later time of the outburst will give the minimum power much less than the actual minimum possible power of that event. 

Also, the minimum power requirement corresponds to the magnetization parameter of $\sigma \sim 1$, while observations show it to be often significantly different. Thus, the actual jet power will be substantially higher than the minimum one. In order to enable a formation of strong hydrodynamic shocks efficiently converting the bulk kinetic energy to nonthermal particles, $\sigma < 1$ is required \citep{ss09,lyubarsky10b,mimica10}, whereas $\sigma > 1$ is required for the efficient magnetic-to-particle energy transfer via magnetic reconnection \citep*{lk01,lyubarsky05,lyubarsky10a,kms13,ss14}. Some ways of obtaining $\sigma\ll 1$ in jets are discussed by \citet*{tmn09} and \citet{komissarov11}. Still, the models of \citet{mzc13} and \citet{z14b} with soft electron spectra and $\gamma_{\min}\sim 1$, which have $\sigma$ much lower than unity may be not realistic. This problem can be resolved if, e.g., the minimum Lorentz factor of the accelerated electrons is assumed to be $\gg 1$, as illustrated by the model 1m by \citet{z14b}, which, at $\gamma_{\rm min}=300$, has $\sigma\la 0.25$.

\section{Conclusions}
\label{conclusions}

We have derived the minimum possible power of jets based from their observed non-thermal synchrotron emission. Our new results give the minimum jet power corresponding to given range of observed synchrotron photon energies, $\epsilon'$. This calculation is qualitatively similar to the well-known one for the minimum energy content, e.g., the minimum power corresponds to an approximate equipartition between the power in particles and in the field. However, it is quantitatively different for jets containing ions. The reason for it is the presence of the ion inertial component in the jet power, which is absent in the energy content. If the electron component dominates, $P_{\rm min}\propto (L'_{\epsilon'})^{4/7} r_{\rm j}^{2/7}$, analogously to the case of the energy content (but taking into account that $P_{\rm j}\propto r_{\rm j}^2$ while $W\propto r_{\rm j}^3$). However, if the ion rest-mass component dominates, we find a remarkable result that $P_{\rm min}\propto (L'_{\epsilon'})^{2/3} r_{\rm j}^{0}$, i.e., the minimum jet power is independent of the radius. This allows for robust estimates of the minimum power even in the absence of information on the spatial extent of the emission. We have also determined the dependencies of the minimum power on the degree of clumpiness.

Our main new results determining the minimum jet power in the case of optically-thin synchrotron emission are given in equations (\ref{Pmin3}--\ref{r_crit}). We also consider cases in which we know the turnover frequency, below which the source becomes optically thick to self-absorption. Our results for that case are given in equations (\ref{a_t}), (\ref{Pminabs}--\ref{Pvar}). If the turnover energy remains not determined, we can use the flux from the partially optically-thick spectral range of extended jets \citep{bk79} to determine the minimum jet power, see equations (\ref{a_extended}--\ref{Pminext}).

Also, the minimum jet power, if dominated by the ionic component, corresponds to the dominance of the magnetic field energy density over that of the electrons, i.e., $\beta^{-1}\gg 1$. Conversely, the electron-$B$ equipartition corresponds in that case to $P_{\rm e}+P_{\rm i}\ll P_B$. Then, the jet power is above the minimum, $P_{\rm j}\gg P_{\rm min}$. This illustrates the difference between the equipartition corresponding to approximately equal energy densities and the equipartition of enthalpies (including the rest mass), i.e., the magnetization parameter of $\sigma\sim 1$, which also approximately corresponds to $P_{\rm e}+P_{\rm i}\sim P_B$.

Taking into account that the usual assumption of spherical (clumpy) source with sharp boundaries is rather unrealistic, we also consider the case of Gaussian overall profiles of the particle and magnetic energy densities, see Appendix \ref{gauss}. The power minimization method can be applied in this case as well. Also, we calculate the corresponding profile of the surface brightness. 

\section*{ACKNOWLEDGMENTS}

I am grateful for Patryk Pjanka, Marek Sikora, {\L}ukasz Stawarz, Markus B{\"o}ttcher and the referee for their valuable comments. This research has been supported in part by the Polish NCN grants 2012/04/M/ST9/00780 and 2013/10/M/ST9/00729.

\appendix

\section{A Gaussian source}
\label{gauss}

An important issue when comparing the above simple theoretical models with observations is determination of the source size. Jets and blobs observed in radio are usually not uniform in brightness, and thus the concept of a source with sharp boundaries is not adequate for them. In most cases, we do not know the actual spatial profiles of the density and magnetic field strength. We may expect, in some models, those distributions peaking at the outer jet boundary. Still, some data show the surface brightness peaking in the centre, e.g., ejections from GRS 1915+105 \citet{mr94}. A natural choice is then the assumption of a Gaussian profile of the density and magnetic field energy density. Namely, we assume the radial distributions of the relativistic electrons and the magnetic energy density to follow
\begin{eqnarray}
\lefteqn{
K(r)=K_0\exp\left[-(r/r_{\rm j})^2\right],\quad B^2(r) =B_0^2\exp\left[-(r/r_{\rm j})^2\right],
\label{K_gauss}}
\end{eqnarray}
where $r_{\rm j}$ is now the average radius, and $K_0$, $B_0$ correspond to $r=0$. We also assume that all other particle densities, e.g., of pairs, follow this dependence. As in the spherical case, we take into account clumping, with the constant local clumping factor, $f$. The average values of $B^2$ and $n$ weighted by the density are $2^{-3/2} B_0^2$ and $2^{-3/2} n_0$. The total number of power-law electrons in the Gaussian sources of both the jet and counterjet is $2\upi^{3/2}n_{\rm pl,0} f r_{\rm j}^3$ (where $n_{\rm pl,0}$ is the central density). This can be multiplied by $\mu_{\rm pl}m_{\rm p}$ to obtain the total mass. The total energy content in both sources is $2\upi^{3/2}u_0 f r_{\rm j}^3$, where $u_0$ is the central energy density in both particles and magnetic field within a clump. Integrating over the jet cross section gives the jet powers as 
\begin{eqnarray}
\lefteqn{P_{\rm e}\simeq 
2\upi  \left(\eta n_{\rm pl,0}\langle\gamma-1\rangle+2 n_{+,0}\right) m_{\rm e}c^3\beta_{\rm j} f (\Gamma_{\rm j} r_{\rm j})^2,\label{pe1}}\\
\lefteqn{P_{\rm i} 
\simeq 2\upi \mu_{\rm pl} n_{\rm pl,0} 
m_{\rm p}c^3 \beta_{\rm j}\Gamma_{\rm j} (\Gamma_{\rm j}-1)f r_{\rm j}^2, \label{pi1}}\\
\lefteqn{P_B=\eta_B (B_0^2/4) \beta_{\rm j} c f(\Gamma_{\rm j} r_{\rm j})^2,
\label{pb1}}
\end{eqnarray}
where $n_{+,0}$ is the central positron density within a clump. We can see that these formulae have the same form as the corresponding ones for spherical sources with the densities and radius equal to the corresponding quantities for the Gaussian case.

The synchrotron luminosity from the source, assumed to have spherical symmetry with the radial distributions given by equation (\ref{K_gauss}), per unit photon energy is (neglecting hereafter effects of time lags and expansion of the sphere),
\begin{equation}
L'_{\epsilon'}\equiv {{\rm d}L'_{\rm S}\over {\rm d}\epsilon'}= {2\upi^{1/2} C_1(p) \sigma_{\rm T}c K_0 B_{\rm cr}^2 f r_{\rm j}^3 \over 3(5+p)^{3/2}}\left(B_0\over B_{\rm cr}\right)^{{p+1}\over 2} 
(\epsilon')^{{1-p}\over 2}.
\label{syn_sphere_gauss}
\end{equation}
The numerical coefficient here is somewhat lower than that for a spherical source characterized by $K_0$, $B_0$ and $r_{\rm j}$. To determine $r_{\rm j}$ from data in such case, the synchrotron emissivity needs to be integrated over the line of sight. This yields the intensity (or surface brightness) profile as a function of the distance, $d'$, from the position of the peak intensity, $I_0$, as
\begin{equation}
I(d')=I_0\, \exp \left[-{5+p\over 4}\left(d'\over r_{\rm j}\right)^2\right].
\label{sphere_profile}
\end{equation}
The intensity drops to $I_0/2$ at
\begin{equation}
d'_{1/2}=r_{\rm j} \sqrt{\ln 16\over 5+p}.
\label{half}
\end{equation}
Representative numerical values are $d'_{1/2}/r_{\rm j}\simeq 0.65$, 0.63, 0.59 for $p=1.5$, 2, 3, respectively. Thus, we see the half-radius is relatively weakly dependent on $p$. Then, we can use the distance at which the intensity decreases by 2 to determine $r_{\rm j}$. The self-absorption optical depth from the source midplane at the distance, $d'$, from the line of sight crossing the centre is
\begin{equation}
\tau_{\rm S}(\epsilon',d')= {\upi^{3\over 2} C_2(p) \sigma_{\rm T}f r_{\rm j} K_0 \epsilon^{-{p+4\over 2}}\over 2(6+p)^{1/2}\alpha_{\rm f}}\left(B_0\over B_{\rm cr}\right)^{p+2\over 2} \!\!
\exp \left[-{6+p\over 4}\left(d'\over r_{\rm j}\right)^2\right],
\label{tau_sphere_gauss}
\end{equation}
which is 1/2 of the total optical depth across the source.

\label{lastpage}


\begin{thebibliography}{}

\bibitem[\protect\citeauthoryear{Atoyan \& Aharonian}{1999}]{aa99} 
Atoyan A.~M., Aharonian F.~A., 1999, MNRAS, 302, 253 

\bibitem[\protect\citeauthoryear{Barniol Duran, Nakar \& Piran}{Barniol Duran et al.}{2013}]{bnp13} 
Barniol Duran R., Nakar E., Piran T., 2013, ApJ, 772, 78 (BNP13)

\bibitem[\protect\citeauthoryear{Blandford \& K{\"o}nigl}{1979}]{bk79} 
Blandford R.~D., K{\"o}nigl A., 1979, ApJ, 232, 34

\bibitem[\protect\citeauthoryear{Bridle}{1967}]{bridle67} 
Bridle A.~H., 1967, Obs, 87, 263 

\bibitem[\protect\citeauthoryear{Brocksopp et al.}{2013}]{brocksopp13} 
Brocksopp C., Corbel S., Tzioumis A., Broderick J.~W., Rodriguez J., Yang J., Fender R.~P., Paragi Z., 2013, MNRAS, 432, 931 

\bibitem[\protect\citeauthoryear{Burbidge}{1956}]{burbidge56} 
Burbidge G.~R., 1956, ApJ, 124, 416 

\bibitem[\protect\citeauthoryear{Chaty, Dubus \& Raichoor}{Chaty et al.}{2011}]{cdr11} 
Chaty S., Dubus G., Raichoor A., 2011, A\&A, 529, A3 

\bibitem[\protect\citeauthoryear{Chevalier}{1998}]{chevalier98} 
Chevalier R.~A., 1998, ApJ, 499, 810 

\bibitem[\protect\citeauthoryear{Dermer \& Atoyan}{2004}]{da04} 
Dermer C.~D., Atoyan A., 2004, ApJ, 611, L9 

\bibitem[\protect\citeauthoryear{Dermer \& Menon}{2009}]{dm09} 
Dermer C.~D., Menon G., 2009, High Energy Radiation from Black Holes: Gamma Rays, Cosmic Rays, and Neutrinos. Princeton Univ.\ Press, Princeton

\bibitem[\protect\citeauthoryear{D{\'{\i}}az Trigo et al.}{2013}]{diaztrigo13} 
D{\'{\i}}az Trigo M., Miller-Jones J.~C.~A., Migliari S., Broderick J.~W., Tzioumis T., 2013, Nat, 504, 260 

\bibitem[\protect\citeauthoryear{Falcke \& Biermann}{1995}]{fb95} 
Falcke H., Biermann P.~L., 1995, A\&A, 293, 665 

\bibitem[\protect\citeauthoryear{Fender}{2006}]{fender06}
Fender, R. P., 2006, in W. H. G. Lewin and M. van der Klis, eds., Compact Stellar X-Ray Sources. Cambridge Univ.\ Press, Cambridge, p.\ 381 

\bibitem[\protect\citeauthoryear{Fender \& Pooley}{2000}]{fp00} 
Fender R.~P., Pooley G.~G., 2000, MNRAS, 318, L1 

\bibitem[\protect\citeauthoryear{Fender et al.}{1999}]{fender99} 
Fender R.~P., Garrington S.~T., McKay D.~J., Muxlow T.~W.~B., Pooley G.~G., Spencer R.~E., Stirling A.~M., Waltman E.~B., 1999, MNRAS, 304, 865

\bibitem[\protect\citeauthoryear{Fender et al.}{2000}]{fender00}
Fender, R. P., Pooley, G. G., Durouchoux, P., Tilanus, R. P. J., Brocksopp, C., 2000, MNRAS, 312, 853

\bibitem[\protect\citeauthoryear{Fender, Belloni \& Gallo}{Fender et al.}{2004}]{fbg04} 
Fender R.~P., Belloni T.~M., Gallo E., 2004, MNRAS, 355, 1105 

\bibitem[\protect\citeauthoryear{Fender, Gallo \& Russell}{Fender et al.}{2010}]{fgr10}
Fender R. P., Gallo E, Russell D. M., 2010, MNRAS, 406, 1425 

\bibitem[\protect\citeauthoryear{Gallo et al.}{2005}]{gallo05} 
Gallo E., Fender R., Kaiser C., Russell D., Morganti R., Oosterloo T., Heinz S., 2005, Nat, 436, 819 

\bibitem[\protect\citeauthoryear{Ghisellini}{1999}]{ghisellini99} 
Ghisellini G., 1999, ASPC, 161, 249 

\bibitem[\protect\citeauthoryear{Ghisellini \& Celotti}{2001}]{gc01} 
Ghisellini G., Celotti A., 2001, MNRAS, 327, 739 

\bibitem[\protect\citeauthoryear{Gleissner et al.}{2004}]{gleissner04} 
Gleissner T., et al., 2004, A\&A, 425, 1061 

\bibitem[\protect\citeauthoryear{Gliozzi, Bodo \& Ghisellini}{Gliozzi et al.}{1999}]{gliozzi99} 
Gliozzi M., Bodo G., Ghisellini G., 1999, MNRAS, 303, L37 

\bibitem[\protect\citeauthoryear{Hornby \& Williams}{1966}]{hw66} 
Hornby J.~M., Williams P.~J.~S., 1966, MNRAS, 131, 237 

\bibitem[\protect\citeauthoryear{Jones, O'Dell \& Stein}{Jones et al.}{1974}]{jos74} 
Jones T.~W., O'Dell S.~L., Stein W.~A., 1974, ApJ, 188, 353 

\bibitem[\protect\citeauthoryear{Kagan, Milosavljevi{\'c} \& Spitkovsky}{Kagan et al.}{2013}]{kms13} 
Kagan D., Milosavljevi{\'c} M., Spitkovsky A., 2013, ApJ, 774, 41 

\bibitem[\protect\citeauthoryear{Komissarov}{2011}]{komissarov11} 
Komissarov S.~S., 2011, MmSAI, 82, 95 

\bibitem[\protect\citeauthoryear{Leahy}{1991}]{leahy91} 
Leahy J.~P., 1991, in P. A. Hughes, ed., Beams and Jets in Astrophysics. Cambridge Univ.\ Press, Cambridge, p.\ 100 

\bibitem[\protect\citeauthoryear{Levinson}{2006}]{levinson06} 
Levinson A., 2006, IJMPA, 21, 6015 

\bibitem[\protect\citeauthoryear{Lind \& Blandford}{1985}]{lb85} 
Lind K.~R., Blandford R.~D., 1985, ApJ, 295, 358 

\bibitem[\protect\citeauthoryear{Longair}{2011}]{longair11} 
Longair M.~S., 2011, High Energy Astrophysics. Cambridge Univ.\ Press, Cambridge

\bibitem[\protect\citeauthoryear{Lyubarsky}{2005}]{lyubarsky05} 
Lyubarsky Y.~E., 2005, MNRAS, 358, 113 

\bibitem[\protect\citeauthoryear{Lyubarsky}{2010a}]{lyubarsky10a} 
Lyubarsky Y., 2010a, ApJ, 725, L234 

\bibitem[\protect\citeauthoryear{Lyubarsky}{2010b}]{lyubarsky10b} 
Lyubarsky Y.~E., 2010b, MNRAS, 402, 353 

\bibitem[\protect\citeauthoryear{Lyubarsky \& Kirk}{2001}]{lk01} 
Lyubarsky Y., Kirk J.~G., 2001, ApJ, 547, 437 

\bibitem[\protect\citeauthoryear{Malyshev, Zdziarski \& Chernyakova}{Malyshev et al.}{2013}]{mzc13} 
Malyshev D., Zdziarski A.~A., Chernyakova M., 2013, MNRAS, 434, 2380

\bibitem[\protect\citeauthoryear{Meier}{2012}]{meier12} 
Meier D., 2012, Black Hole Astrophysics: The Engine Paradigm. Berlin: Springer

\bibitem[\protect\citeauthoryear{Miller-Jones, Fender \& Nakar}{Miller-Jones et al.}{2006}]{millerjones06} 
Miller-Jones J.~C.~A., Fender R.~P., Nakar E., 2006, MNRAS, 367, 1432 

\bibitem[\protect\citeauthoryear{Mimica \& Aloy}{2010}]{mimica10} 
Mimica P., Aloy M.~A., 2010, MNRAS, 401, 525 

\bibitem[\protect\citeauthoryear{Mirabel \& Rodr{\'{\i}}guez}{1994}]{mr94} 
Mirabel I.~F., Rodr{\'{\i}}guez L.~F., 1994, Nat, 371, 46

\bibitem[\protect\citeauthoryear{Narayan \& McClintock}{2012}]{nm12} 
Narayan R., McClintock J.~E., 2012, MNRAS, 419, L69 

\bibitem[\protect\citeauthoryear{Orosz et al.}{2011}]{orosz11} 
Orosz J.~A., McClintock J.~E., Aufdenberg J.~P., Remillard R.~A., Reid 
M.~J., Narayan R., Gou L., 2011, ApJ, 742, 84

\bibitem[\protect\citeauthoryear{Pacholczyk}{1970}]{pacholczyk70} 
Pacholczyk A.~G., 1970, Radio astrophysics. Nonthermal processes in galactic and extragalactic sources. San Francisco: Freeman  

\bibitem[\protect\citeauthoryear{Reid et al.}{2011}]{reid11} 
Reid M.~J., McClintock J.~E., Narayan R., Gou L., Remillard R.~A., Orosz 
J.~A., 2011, ApJ, 742, 83 

\bibitem[\protect\citeauthoryear{Russell, Gallo \& Fender}{Russell et al.}{2013}]{rgf13} 
Russell D.~M., Gallo E., Fender R.~P., 2013, MNRAS, 431, 405 

\bibitem[\protect\citeauthoryear{Scheuer \& Williams}{1968}]{sw68} 
Scheuer P.~A.~G., Williams P.~J.~S., 1968, ARA\&A, 6, 321 

\bibitem[\protect\citeauthoryear{Scott \& Readhead}{1977}]{sr77} 
Scott M.~A., Readhead A.~C.~S., 1977, MNRAS, 180, 539 

\bibitem[\protect\citeauthoryear{Sikora et al.}{1997}]{sikora97}
Sikora M., Madejski G., Moderski R., Poutanen J., 1997, ApJ, 484, 108

\bibitem[\protect\citeauthoryear{Sironi \& Spitkovsky}{2009}]{ss09} 
Sironi L., Spitkovsky A., 2009, ApJ, 698, 1523 

\bibitem[\protect\citeauthoryear{Sironi \& Spitkovsky}{2014}]{ss14} 
Sironi L., Spitkovsky A., 2014, ApJ, 783, L21 

\bibitem[\protect\citeauthoryear{Slish}{1963}]{slish63} 
Slish V.~I., 1963, Nat, 199, 682 

\bibitem[\protect\citeauthoryear{Stawarz et al.}{2004}]{stawarz04} 
Stawarz {\L}., Sikora M., Ostrowski M., Begelman M.~C., 2004, ApJ, 608, 95 

\bibitem[\protect\citeauthoryear{Stawarz et al.}{2013}]{stawarz13} 
Stawarz {\L}., et al., 2013, ApJ, 766, 48 

\bibitem[\protect\citeauthoryear{Steiner, McClintock \& Narayan}{Steiner et al.}{2013}]{smn13} 
Steiner J.~F., McClintock J.~E., Narayan R., 2013, ApJ, 762, 104 

\bibitem[\protect\citeauthoryear{Stirling et al.}{2001}]{stirling01} 
Stirling A.~M., Spencer R.~E., de la Force C.~J., Garrett M.~A., Fender R.~P., Ogley R.~N., 2001, MNRAS, 327, 1273 

\bibitem[\protect\citeauthoryear{Tchekhovskoy, McKinney \& Narayan}{Tchekhovskoy et al.}{2009}]{tmn09} 
Tchekhovskoy A., McKinney J.~C., Narayan R., 2009, ApJ, 699, 1789 

\bibitem[\protect\citeauthoryear{Williams}{1963}]{williams63} 
Williams P.~J.~S., 1963, Nat, 200, 56 

\bibitem[\protect\citeauthoryear{Willott et al.}{1999}]{willott99} 
Willott C.~J., Rawlings S., Blundell K.~M., Lacy M., 1999, MNRAS, 309, 1017 

\bibitem[\protect\citeauthoryear{Zdziarski}{2014}]{zdz14} 
Zdziarski A.~A., 2014, MNRAS, 444, 1113

\bibitem[\protect\citeauthoryear{Zdziarski, Lubi{\'n}ski \& Sikora}{Zdziarski et al.}{2012}]{zls12} 
Zdziarski A.~A., Lubi{\'n}ski P., Sikora M., 2012, MNRAS, 423, 663

\bibitem[\protect\citeauthoryear{Zdziarski et al.}{2014a}]{z14a} 
Zdziarski A.~A., Stawarz {\L}., Pjanka P., Sikora M., 2014a, MNRAS, 440, 2238 

\bibitem[\protect\citeauthoryear{Zdziarski et al.}{2014b}]{z14b} 
Zdziarski A.~A., Pjanka P., Sikora M., Stawarz {\L}., 2014b, MNRAS, 442, 3243

\end{thebibliography}
\end{document}